\documentclass[sigconf]{acmart}
\acmConference[ICSE 2022]{The 44th International Conference on Software Engineering}{May 21–29, 2022}{Pittsburgh, PA, USA}
\AtBeginDocument{%
  \providecommand\BibTeX{{%
    \normalfont B\kern-0.5em{\scshape i\kern-0.25em b}\kern-0.8em\TeX}}}

\copyrightyear{2022} 
\acmYear{2022} 
\setcopyright{acmlicensed}\acmConference[ICSE '22]{44th International Conference on Software Engineering}{May 21--29, 2022}{Pittsburgh, PA, USA}
\acmBooktitle{44th International Conference on Software Engineering (ICSE '22), May 21--29, 2022, Pittsburgh, PA, USA}
\acmPrice{15.00}
\acmDOI{10.1145/3510003.3510153}
\acmISBN{978-1-4503-9221-1/22/05}

\usepackage{amsmath}
\usepackage[frozencache,cachedir=.]{minted}

\begin{document}

\title{Learning to Reduce False Positives in Analytic Bug Detectors}

\author{Anant Kharkar}
\affiliation{
  \institution{Microsoft}
  \city{Redmond}
  \state{Washington}
  \country{USA}
}

\author{Roshanak Zilouchian Moghaddam}
\affiliation{
  \institution{Microsoft}
  \city{Redmond}
  \state{Washington}
  \country{USA}
}

\author{Matthew Jin}
\affiliation{
  \institution{Microsoft}
  \city{Redmond}
  \state{Washington}
  \country{USA}
}

\author{Xiaoyu Liu}
\affiliation{
  \institution{Microsoft}
  \city{Redmond}
  \state{Washington}
  \country{USA}
}

\author{Xin Shi}
\affiliation{
  \institution{Microsoft}
  \city{Redmond}
  \state{Washington}
  \country{USA}
}

\author{Colin Clement}
\affiliation{
  \institution{Microsoft}
  \city{Redmond}
  \state{Washington}
  \country{USA}
}

\author{Neel Sundaresan}
\affiliation{
  \institution{Microsoft}
  \city{Redmond}
  \state{Washington}
  \country{USA}
}


\begin{abstract}
Due to increasingly complex software design and rapid iterative development, code defects and security vulnerabilities are prevalent in modern software. In response, programmers rely on static analysis tools to regularly scan their codebases and find potential bugs. In order to maximize coverage, however, these tools generally tend to report a significant number of false positives, requiring developers to manually verify each warning. To address this problem, we propose a Transformer-based learning approach to identify false positive bug warnings. We demonstrate that our models can improve the precision of static analysis by 17.5\%. In addition, we validated the generalizability of this approach across two major bug types: null dereference and resource leak.

\end{abstract}



\maketitle

\section{Introduction}

Software defects (bugs) that go undetected during the development process can cause software failure, resulting in financial and reputational harm to companies and a host of problems for users of buggy software. Developers often rely on static analysis tools to scan their codebases and find potential bugs. Despite their benefits, static analysis tools are not consistently used in many software projects \cite{Ayewah-MS2008}. Previous work has attributed their inconsistent usage to high false positive rates and ineffective presentation of warnings \cite{johnson-icse13}.

Developing any static analyzer is a non-trivial task due to the trade-off between precision and recall; it is challenging to report only correct bugs (precision) while covering all bugs with a similar pattern (coverage/recall). Balancing these two objectives manually is difficult and can result in analyzers with high false positive rate (low precision). Analyzers with high initial precision can also degrade in predictive performance as the nature of bugs changes over time. Continuously updating and maintaining static analyzers to handle concept drift can be costly \cite{bielik2017learning}.

Previous research has investigated various methods to improve static analysis false positive rates. In particular, researchers have explored eliminating bugs along infeasible paths using syntactic model-checking \cite{junker2012smt}; eliminating all the bugs that are similar to a false positive based on similarity of modification points \cite{muske2013review}; and using a two-staged error ranking strategy where false positive patterns are learned after manual labeling in the first stage \cite{shen-icst2011, Tripp-CCS2014}. Our work uniquely contributes to this line of prior work by leveraging state-of-the-art neural models to automatically refine the output of static analyzers. 

Beyond traditional rule-based tools, there has been significant recent work leveraging machine learning for software bug and vulnerability detection in various languages, including C/C++ \cite{russell2018automated, li2018vuldeepecker}, Java \cite{pang2015predicting}, and JavaScript \cite{pradel-oopsla-2018}. However, much like rule-based analyzers, these machine learning models often suffer from low precision when applied in real world settings. Another challenge with some machine learning approaches is the need to develop new models to capture new types of vulnerabilities. Unlike this line of work, we do not use machine learning to detect bugs directly. Instead, we leverage machine learning to augment existing static analyzers. We believe this strategy yields the best of both worlds, where machine learning complements the capabilities of current static analyzers. 

To augment static analyzers, we explored several models, including a feature based model and two neural models. Our feature-based model includes a set of carefully handcrafted features extracted from source code. Our neural models were inspired by the recent successes of transformer models in code search and document generation \cite{feng2020codebert} as well as code completion \cite{Svyatkovskiy-corr-2020}. One of our neural models learns from labeled data (DeepInferEnhance), while the other is applied in a zero-shot setting without the need for further training or finetuning (GPT-C \cite{Svyatkovskiy-corr-2020}). We conducted an experiment with all the models on bugs identified by Infer, an interprocedural static analyzer that detects bugs in Java, C++, and C\#. Our results show that we can improve the precision of Infer's analysis by up to 17\%.

\section{Related Work}
We describe the prior work on static analyzers and the use of machine learning for bug detection.

\subsection{Static Analysis-Based Bug Detection}
Rule-based systems and static analyzers have been widely adopted for detecting software bugs \cite{xu2010memory, viega2000its4, sonarqube, coverity}. However, one of the barriers to consistent usage of static analyzers is their high false positive rate \cite{johnson-icse13}. Previous work has explored various ways of reducing this false positive rate. For instance, Junker et al. \cite{junker2012smt} leveraged syntactic model-checking to eliminate infeasible paths (program slices). An implementation of their approach on Goanna, an static analyzer for C/C++ programs showed that they could exclude the majority of false positives. Muske, et al. \cite{muske2013review} implemented a partitioning mechanism to partition similar warnings based on the modified variables and modification points. A whole partition is then considered false positive once its leader is determined as false positive. Shen et al. \cite{shen-icst2011} developed EFindBugs, which uses a two-staged error ranking strategy to deal with the false positives issue in FindBugs \cite{Ayewah-oopsla2007}. EFindBugs first reports warnings on a sample program. Once the warnings are manually labeled, the tool learns what bug patterns to eliminate on the second run against the user application. Similarly, ALETHEIA learns users preferences from manual labeling on a smaller set \cite{Tripp-CCS2014}. Our work uniquely contributes to this line of prior work by exploring the use of state-of-the-art neural models to automatically refine the output of static analyzers by removing false positives.

\subsection{Learning-Based Bug Detection}
Beyond rule-based tools, there has been significant recent work on data-driven and machine learning approaches to detect software bugs and vulnerabilities. For instance, Russell et al. \cite{russell2018automated} proposed a machine learning method for detecting software vulnerabilities in C/C++ code bases. Similarly, Choi et al. \cite{choi-corr-2017} trained a memory neural network to detect a variety of buffer overruns in C-style code. Li et al. \cite{li2018vuldeepecker} trained a recurrent neural network (RNN) to detect two specific types of vulnerabilities related to improper use of library/API functions. Bugram \cite{wang2016bugram} leveraged n-gram language models to identify low probability token sequences in code as bugs. Pang et al. \cite{pang2015predicting} trained a machine learning model to predict static analyzer labels for Java source code. Finally, DeepBugs \cite{pradel-oopsla-2018} trained a classifier that distinguishes correct from incorrect code for three classes of bugs (swapped function arguments, wrong binary operator, and wrong operand in a binary operation) in JavaScript. However, the majority of machine learning solutions suffer from low precision when applied on real world settings. Another challenge with some machine learning approaches is the need to develop new models to capture new types of vulnerabilities. By leveraging machine learning to augment existing static analyzers, our work creates the best of both worlds, where machine learning will complement the capabilities of current static analyzers to generate more precise results.  

\begin{figure}[h!]
    \begin{minted}[fontsize=\footnotesize, linenos, firstnumber=8]{csharp}
    static void Main(string[] args)
    {
        var returnNull = ReturnNull();
        _ = returnNull.Value;
    }

    private static NullObj ReturnNull()
    {
        return null;
    }

    internal class NullObj
    {
        internal string Value { get; set; }
    }
    \end{minted}
    \mintinline{vim}{/Examples/NullDeref/Program.cs:11}
    \mintinline{vim}{error: NULL_DEREFERENCE} pointer 'returnNull' could be null and is dereferenced at line 11, column 13.
    \caption{An example of a Null Dereference detected by Infer}
    \label{fig:Infer_example}
\end{figure}

\section{Infer}
Our false positive reduction approach can work with any static analyzer for which labeled data is available. Our experiments specifically targeted Infer, an interprocedural static analyzer that is used to detect a variety of bugs in Java and C++. The recent release of Infer\# also added support for bug detection in C\# projects. Infer uses separation logic, a program logic for reasoning about memory manipulations, to prove certain memory safety conditions and create program state summaries for each method in a code base. 
Infer's analysis examines multiple methods in order to identify bugs in code. When analyzing each method, Infer formulates pre- and post-conditions that describe the impact of the method on the memory state of the program. When analyzing a method invocation, Infer uses the conditions of the callee to form logical predicates for the caller. Thus, Infer analyzes the entire call stack of a program by composing logical predicates from all nested callee methods. Figure \ref{fig:Infer_example} shows an example of a null dereference bug identified by Infer. 

\begin{figure*}[ht!]
\begin{minted}[fontsize=\footnotesize, linenos, firstnumber=136]{java}
private void dumpLog(File logFile, long startOffset, long endOffset, ArrayList<String> blobs) throws IOException {
    Map<String, LogBlobStatus> blobIdToLogRecord = new HashMap<>();
    final Timer.Context context = metrics.dumpLogTimeMs.time();
    try {
      dumpLog(logFile, startOffset, endOffset, blobs, blobIdToLogRecord);
      long totalInConsistentBlobs = 0;
      for (String blobId : blobIdToLogRecord.keySet()) {
        LogBlobStatus logBlobStatus = blobIdToLogRecord.get(blobId);
        if (!logBlobStatus.isConsistent) {
          totalInConsistentBlobs++;
          logger.error("Inconsistent blob " + blobId + " " + logBlobStatus);
        }
\end{minted}
\mintinline{vim}{ambry-tools/src/main/java/com.github.ambry/store/DumpLogTool.java:144}

\mintinline{vim}{error: NULL_DEREFERENCE} object `logBlobStatus` last assigned on line 143 could be null and is dereferenced at line 144.
\caption{An example of a false positive warning from Infer. Infer warns that \mintinline{java}{logBlobStatus} can be null. This occurs if \mintinline{java}{blobId} is not a valid key of \mintinline{java}{blobIdToLogRecord}. The warning is incorrect, since \mintinline{java}{blobId} comes from \mintinline{java}{blobIdToLogRecord}'s key set.}
\label{figure:false_positive}
\end{figure*}

We decided to focus on Infer for three reasons. First, unlike the majority of common analyzers that only consider the context of a single method (i.e. intraprocedural), Infer's analysis is interprocedural and its context can stretch across several methods. Second, as opposed to many analyzers that rely on developer annotations to detect certain bugs, Infer's analysis is automated and does not rely on annotations. Finally, due to incremental change analysis, Infer can scale well on large production codebases.

Like other static analyzers, Infer is also prone to false positives. Figure \ref{figure:false_positive} shows an example, in which Infer reports that the variable \mintinline{java}{logBlobStatus} can be null, since it is assigned by calling \mintinline{java}{get()} on a map; if the key is not present in the map, \mintinline{java}{get()} will return null. 
However, the key \mintinline{java}{blobId} comes from the \mintinline{java}{keySet} of the same map, meaning the value must exist in the map and \mintinline{java}{logBlobStatus} cannot be null. Infer is not able to recognize the coding convention of iterating over a map's key set and incorrectly triggers a null dereference warning. Language models, which are trained to identify patterns across a large corpus of code, can recognize such idioms. This motivated us to turn to machine learning to detect false positives reported by Infer. Indeed, our model identifies this specific warning as a false positive.

\section{False Positive Reduction}
False positive Infer warnings share common characteristics and follow patterns in coding conventions, as described in the example above. This motivated us to turn to machine learning as a means of capturing these patterns and identifying false positive warnings. We experimented with several models, including a feature-based model and two transformer-based neural models. Below, we present the data set we used for training and testing these models, as well as the details of each model.

\subsection{Data Collection}
Our data set consists of 539 \emph{null dereference} warnings generated by running Infer on seven Java repositories. Null dereference bugs occur when a pointer that can potentially be null is dereferenced. To create a diverse dataset, we chose several open source projects and two proprietary projects. The projects include back-end service components (\emph{Ambry}, \emph{Azure SDK}, and \emph{Nacos}), build plugins (\emph{Azure Maven Plugins}), and browser automation (\emph{Playwright}). Table \ref{table:repo-stats} summarizes the projects in our dataset, including the number of total warnings and true positives reported by vanilla Infer. Project A and Project B denote the proprietary projects.

\begin{table}[h!]
\centering 
 \caption{Summary statistics of null dereference warnings. Total warnings and true positives are as reported by vanilla Infer.}
 \label{table:repo-stats}
\begin{tabular}{c c c c c} 
 \hline
 Name       & Lines of  & Total    & True      & Precision \\ 
            & Code      & Warnings & Positives \\
 \hline\hline
Project A	&    35,527 &       57 &	    47 &    82.4\% \\\hline
Project B   &	 66,346 &       33 &        30 &    90.9\% \\\hline
Ambry       &   138,947 &       25 &        17 &    68.0\% \\\hline
Azure SDK   & 3,555,286 &      343 &       272 &    79.3\% \\\hline
Playwright  &	 21,094 &       18 &         3 &    16.7\% \\\hline
Nacos       &	 62,443 &       37 &        13 &    35.1\% \\\hline
Azure Maven &	 23,995 &       26 &        10 &    38.5\% \\
Plugins\\\hline
\textbf{Total} & \textbf{3,903,638} & \textbf{539} & \textbf{392} & \textbf{72.7\%}\\\hline
\end{tabular}
\end{table}
 
Each warning was investigated and labeled as valid (true positive) or invalid (false positive) by experienced developers. The need for manual labeling presents a bottleneck to scaling up to larger data sets. In the end, the developers identified 392 of the warnings as true positives (72.7\% precision) and 147 as false positives. Precision for individual repositories varied between 16\% to 90\%; the lower end of this range can correspond to poor experience for developers of those projects. There is significant opportunity for machine learning to benefit the experience by improving precision.

For each warning, we record the following information:
\begin{itemize}
    \item the \emph{label}: whether the warning was legitimate or not
    \item the \emph{location} of the warning: includes the file name and line number where the warning occurred
    \item the \emph{code}: this is the code snippet on the line of warning
    \item the error \emph{message}: the error message produced by Infer
    \item the \emph{local context} around the warning: consists of all lines of code from beginning of the surrounding function to the line of the warning.
    \item the \emph{non-local context}: includes the content of functions that were called in the current context.
\end{itemize}

\begin{figure*}[ht!]
\begin{minted}[fontsize=\footnotesize, linenos, firstnumber=461]{java}
Datacenter datacenterToAdd = hardwareLayout.findDatacenter(dataCenterName);
List<Disk> disksForReplicas =
    allocateDisksForPartition(numberOfReplicasPerDatacenter, capacityOfReplicasInBytes, datacenterToAdd,
        attemptNonRackAwareOnFailure);
partitionLayout.addNewReplicas((Partition) partitionId, disksForReplicas);
System.out.println("Added partition " + partitionId + " to datacenter " + dataCenterName);
\end{minted}
\rule{\textwidth}{0.4pt}
\begin{minted}[fontsize=\footnotesize, linenos, firstnumber=198]{java}
public Datacenter findDatacenter(String datacenterName) {
    for (Datacenter datacenter : datacenters) {
      if (datacenter.getName().compareToIgnoreCase(datacenterName) == 0) {
        return datacenter;
      }
    }
    return null;
}
\end{minted}
\mintinline{vim}{ambry-clustermap/src/main/java/com.github.ambry.clustermap/StaticClusterManager.java:463}

\mintinline{vim}{error: NULL_DEREFERENCE} object `datacenterToAdd' last assigned on line 461 could be null and is dereferenced by call to `allocateDisksForPartition(...)' at line 463.
\caption{An example of an interprocedural bug detected by Infer. Infer reports that \mintinline{java}{datacenterToAdd} (top, line 461) can be null. To determine if this is the case, an investigator must find the implementation of \mintinline{java}{findDataCenter()} (bottom), which is used to assign the value of \mintinline{java}{datacenterToAdd}. Since \mintinline{java}{findDataCenter()} explicitly contains the line \mintinline{java}{return null}, \mintinline{java}{datacenterToAdd} can be null and the warning is reasonable.}
\label{figure:nonlocal}
\end{figure*}

The \emph{non-local context} enables us to account for the interprocedural nature of Infer. 
To obtain interprocedural information, we collect and use the content of certain methods invoked in the local context that can impact the value of the null pointer. For example, for some null dereference warnings, the null pointer originates as the return value of a method; we retrieve the body of this method as non-local context.

Figure \ref{figure:nonlocal} demonstrates the importance of non-local context. Infer reports that the variable \mintinline{java}{datacenterToAdd} assigned on line 461 (top) can be null. To investigate this, a developer must look into the \mintinline{java}{findDatacenter} method in a different file (bottom). Here, we can see that \mintinline{java}{findDatacenter} can return null on line 204 if none of the \mintinline{java}{datacenters} match the argument, meaning it is possible for \mintinline{java}{datacenterToAdd} to be null when it is dereferenced. Therefore, in order to determine whether this warning is correct, the content of the callee method (\mintinline{java}{findDatacenter}) is necessary. Although it is possible for a null pointer to originate from multiple nested method calls, we found that in most cases, collecting the immediate callee was sufficient.

\subsection{Feature-Based False Positive Reduction}
As a baseline, we extracted feature vectors from our data and trained a classifier to predict whether a warning is a false positive. Our features included: 
\begin{itemize}
\item whether the non-local context explicitly contains the line \linebreak \mintinline{java}{return null;}. If this line exists in non-local context, then it is possible for the callee to return null, and the  variable that holds the return value in the caller can be null.

\item whether a null-check method appears in the context of the warning. For some Infer warnings, the dereferenced variable is verified to be non-null earlier in the method using special null-check methods (e.g. \mintinline{java}{Objects.requireNonNull()}). Since these null-check methods belong to external libraries, Infer is unable to understand their behavior, resulting in false positive warnings.

\item whether a dereferenced variable is a class field. In practice, Infer's logic makes errors when tracking the state of class fields and often incorrectly treats them as nullable.

\item whether an implicit cast of a wrapper class to a primitive type occurs on the warning line. In our analysis, we realized that implicit casts can be the cause of many null pointer issues. For example, when the code includes a map object with primitive-type values (e.g. HashMap with double values), the map's values must instead be wrapper class objects (Double) instead of primitives (double), since maps in Java cannot take primitives. Values retrieved from the map are often stored in primitive-type variables, causing an implicit cast (see Figure \ref{fig:implicit_cast}). If the wrapper object is null, this cast operation causes a null dereference.
\end{itemize}

We trained a logistic regression classifier on these features. Since the limited size of our data prevents us from using a simple train-test split, we instead used 5-fold cross-validation for training and evaluation. In a realistic scenario, the model would not have access to training data from the same project for which it is making predictions. However, as shown in Table \ref{table:repo-stats}, the projects that comprise our dataset vary widely in the number of warnings, and attempting to separate repos across different folds would result in insufficient training data in several folds. To partially mitigate this issue, we ensure that all warnings from the same file appear in the same fold.

\begin{figure}[ht!] 
\begin{minted}[fontsize=\footnotesize, linenos, ]{java}
HashMap<String, Double> m = new HashMap<String, Double>();
m.put("Bla", new Double(1.0));
//below line will cause an implicit cast operation
double v = m.get("Bla");
    \end{minted}
    
    \caption{Example of an implicit cast of a wrapper class (Double) to primitive type (double) in line 4. Calling \mintinline{java}{get()} on the map \mintinline{java}{m} returns a Double, which is implicitly cast to double to comply with the type of \mintinline{java}{v}.}
    \label{fig:implicit_cast}
\end{figure}

\subsection{Neural False Positive Reduction}
Engineered features, while easy to understand, are inflexible and cannot automatically learn new patterns from data. Deep learning models, particularly transformers, are able to better capture the complexities of modern source code. Transformers are deep neural networks that leverage attention mechanisms to learn patterns in sequential data, such as language. They contain billions of parameters and can leverage massive datasets to learn representations of language patterns. They have achieved state-of-the-art results for applications in natural language processing (NLP) such as machine translation, question answering, and document summarization \cite{vaswani2017attention}. Transformers are usually pretrained on a large unlabeled corpus and further finetuned on task-specific labeled data to perform classification or language generation.

The wealth of open source code available on GitHub has inspired researchers to train a variety of transformers on open source code and finetune them to support many downstream tasks such as code completion\cite{Svyatkovskiy-corr-2020}, documentation generation\cite{feng2020codebert}, automated code review \cite{tufano2021autocodereview}, software traceability \cite{lin2021traceability}, and code search using natural language. 

Two major categories of transformers are auto-generative models and auto-regressive models. Auto-generative models such as BERT \cite{bert} are trained to reproduce their inputs, while auto-regressive model produce the next token in the sequence. In this work, we leverage two transformers to help reduce Infer's false positive rate. The first model, DeepInferEnhance, is a customized version of CodeBERTa: an auto-generative encoder model similar to RoBERTa \cite{liu2019roberta}. The second model is GPT-C, an auto-regressive model with only decoder layers (similar to GPT-2 \cite{radford2019language} and GPT-3 \cite{gpt3}). Both models require only source code as input, rather than any intermediate code structure such as syntax trees or control flow graphs.

\subsubsection{DeepInferEnhance}
CodeBERTa is a pre-trained transformer based on the RoBERTa \cite{liu2019roberta} architecture and developed by HuggingFace \cite{wolf2020huggingfaces}. The model was pre-trained on CodeSearchNet \cite{husain2020codesearchnet}: a multilingual source code corpus of 2 million functions (with comments and docstrings) from GitHub. CodeSearchNet consists of functions from Go, Java, JavaScript, PHP, Python, and Ruby. CodeBERTa was inspired by the success of CodeBERT \cite{feng2020codebert}, an application of the BERT architecture to source code. CodeBERT was also trained on CodeSearchNet and yielded state-of-the-art results for tasks such as code search and documentation generation. Furthermore, CodeBERT's promising results in zero-shot settings showed the power of its representations.

We decided to use the more lightweight and efficient CodeBERTa architecture. However, we were interested in applying Infer to both Java and C\# code, and C\# was notably absent from CodeBERTa's training dataset. Therefore, we pretrained an identical CodeBERTa model on a corpus of 2 million Java and C\# functions that we collected from GitHub. Like the original CodeBERTa, our model is pretrained using a masked language modeling (MLM) objective. 

Encoder-based transformers like CodeBERTa, which incorporate information from both sides of the current position, can learn to create efficient representations of their entire input. Through transfer learning, these representations can then be used to solve more specific tasks. We sought to transfer our pretrained model's learned representations to the task of identifying false positive Infer warnings. Therefore, we added a sequence classification head to this model in order to classify warnings as true positive or false positive. We finetuned the model on our dataset of Infer warnings by freezing all layers except for the classification head. The inputs for finetuning are strings of code context, and the labels are boolean indicators of valid or invalid warnings. Our final model consists of a 6-layer encoder and 2-layer classification head, with a total of 83 million parameters. We call this model \emph{DeepInferEnhance}.
\\\\
\subsubsection{GPT-C}
Unlike auto-generative models, which learn representations to reproduce their input, auto-regressive (generative) models learn to create new text. GPT-3 is one example of such a generative model \cite{gpt3}. Because of the scarcity of labeled Infer warnings for supervised learning, we turned to generative models and used code completion recommendations as a signal of the legitimacy of Infer warnings. Many null dereference warnings can be resolved - even if sub-optimally - by introducing a null check before the dereference. Similarly, many resource leak bugs can be fixed by explicitly releasing the leaked resource. If a generative model recommends a null check or resource release, this may indicate that the corresponding warning is indeed legitimate, since the model deemed that such a fix is necessary. Our intuition is that the model may have a fuzzy understanding that a null check or resource release is required. 

To generate these code recommendations, we use GPT-C \cite{Svyatkovskiy-corr-2020}, a generative transformer based on the well-known GPT-2 \cite{radford2019language}. This model was designed and trained for code line completion and represents the state of the art in this field; it was implemented as part of the IntelliCode Compose web service. This model is also multilingual and was pretrained on C\#, Python, C++, Java, JavaScript, TypeScript, Go, PHP, Ruby, and C. GPT-C takes in a partially written method body and uses multi-headed self-attention to predict the next line. We use it in a zero-shot setting: unlike DeepInferEnhance, we do not train GPT-C ourselves, but instead rely on its pretrained parameters. We do not verify the syntactic correctness of the generated code, but rather use it as a “fuzzy” signal only to determine if a warning is valid or not. For null dereference warnings, if the model generates a null check statement at a line before a null pointer warning occurs, we consider that warning valid.

GPT-C is trained specifically for line completion, rather than whole line generation. This means that, rather than generating a full line of code from previous lines, GPT-C expects an incomplete line of code at the end of its input and generates code to complete this line. This incomplete trailing line is a \emph{prompt} and consists of several tokens at the beginning of the final line. For our objective of predicting Infer warning validity, we provide specific prompts to GPT-C for each warning type. For null dereferences, the input prompts are the prefixes of 7 different null check statements (e.g. \mintinline{java}{if} or \mintinline{java}{Debug.Assert}). For each prompt, we use GPT-C with beam search to generate line completion recommendations. With a beam size of 5, this results in 35 recommendations per warning. We also prepend non-local context to the input where possible. Figure \ref{figure:gptc_success} shows an example input. 

\begin{figure*}[ht!]

\begin{minted}[fontsize=\footnotesize, linenos, firstnumber=201]{java}
for (Node dataNode : nodes) {
    if (allocatedDisks.size() == numberOfReplicas) {
      break;
    }
    Disk disk = dataNode.getDiskWithMostCapacity(replicaSize);
    allocatedDisks.add(disk);
    disk.freeCapacity = disk.freeCapacity - replicaSize;
\end{minted}
\rule{\textwidth}{0.4pt}
\begin{minted}[fontsize=\footnotesize]{java}
public Disk getDiskWithMostCapacity(long replicaSize) {
    Disk minDisk = null;
    for (Disk disk : disks) {
      if ((minDisk == null || minDisk.freeCapacity < disk.freeCapacity) && disk.freeCapacity >= replicaSize) {
        minDisk = disk;
      }
    }
    return minDisk;
}

public static void Strategy3(Datacenter dc, List<Partition> partitions, int numberOfPartitions, int numberOfReplicas,
      long replicaSize) {
    for (int i = 0; i < numberOfPartitions; i++) {
      List<Node> nodes = dc.nodes;
      Collections.shuffle(nodes);
      List<Disk> allocatedDisks = new ArrayList<Disk>();

      for (Node dataNode : nodes) {
        if (allocatedDisks.size() == numberOfReplicas) {
          break;
        }
        Disk disk = dataNode.getDiskWithMostCapacity(replicaSize);
        allocatedDisks.add(disk);
        if (
\end{minted}
\caption{An example of a legitimate Infer warning (top) and the corresponding input to GPT-C (bottom). Infer reports that \mintinline{java}{disk}, which is assigned by \mintinline{java}{getDiskWithMostCapacity()} (bottom) can be null and is dereferenced on line 207. The GPT-C input is constructed by appending a prompt to the method body preceding this line, as well as prepending the non-local context method \mintinline{java}{getDiskWithMostCapacity()}. Here GPT-C correctly predicts a null check and therefore this warning is regarded as legitimate by the GPT-C based model.}
\label{figure:gptc_success}
\end{figure*}

Each Infer warning has an associated file path and line number that correspond to the method where the warning occurs; we call this the target method. For both transformer models, the input to the neural network includes the source code of the target method up to (but not including) the line of the warning. For GPT-C, we include two additional components: the non-local context method body preceding the target method and a line completion prompt immediately following the target method.

\break
\section{Experiments}
We performed two experiments to better understand how these models perform in a real-world setting. The first experiment, summarized in Table \ref{table:null_deref_results}, was focused on comparing effectiveness of our feature-based and neural models. The second experiment focused on verifying the generalizability of our neural approaches when applied to a different bug type. Since our objective is to eliminate false positives reported by Infer, our primary metric to evaluate our models is the relative precision improvement over vanilla Infer. We also measure recall with respect to Infer's true positive warnings: a recall of 100\% means that all of the true positive warnings from vanilla Infer were reported. By construction, none of the approaches in this work report new warnings beyond those originally reported by Infer.

\begin{table}[h!]
\centering 
 \caption{Performance of machine learning for removing false positive null dereference warnings}
 \label{table:null_deref_results}
\begin{tabular}{c c c c c} \hline
Approach            & Precision & $\Delta$ Precision & Recall \\
\hline\hline
Baseline            &    72.7\% &                  - &  100\% \\\hline
Feature-Based       &    78.7\% &            +8.26\% & 65.1\% \\
DeepInferEnhance    &    83.7\% &           +15.13\% & 88.3\% \\
GPT-C               &    85.4\% &           +17.47\% & 83.7\% \\
\hline
\end{tabular}
\end{table}

\subsection{Experiment 1: Comparing feature-based and neural models}

\subsubsection{Feature-Based Model}
The simplest data-driven approach to identify false positive Infer warnings is to manually search for patterns in the warnings. The handcrafted features we collected for our logistic regression model capture the patterns that we discovered from manual review of our dataset. This feature-based model was able to improve Infer's precision by 8\%, but with significant reduction in recall. Since source code can be inherently complex, it is unsurprising that simple handcrafted features are insufficient to identify false positive warnings. 

\subsubsection{DeepInferEnhance}
Since handcrafted features cannot adequately represent source code, we turned to deep learning to automatically learn patterns in code that indicate the legitimacy of Infer warnings. We took a traditional supervised transfer learning approach, using our dataset of labeled warnings to finetune our DeepInferEnhance model. The results show that this model greatly improved precision and recall compared to the feature-based model. 

Transformers are generally finetuned on much larger datasets than the several hundred warnings we used. However, DeepInferEnhance was still able to learn patterns that provided a significant improvement in precision. One such pattern occurs when \mintinline{java}{null} is explicitly passed as an argument to a method. Even if the method handles null arguments, Infer still reports a null dereference warning, which is often a false positive. DeepInferEnhance is able to learn this pattern purely from the code itself.

\begin{figure*}[h]
\begin{minted}[fontsize=\footnotesize, linenos, firstnumber=17]{java}
public class DefaultAzureMessageHandler implements AzureMessageHandler {

    @Nullable
    private InvocableHandlerMethod handlerMethod;

    private Class<?> messagePayloadType;
    
    private String createMessagingErrorMessage(String description) {
        InvocableHandlerMethod handlerMethod = getHandlerMethod();
        StringBuilder sb =
                new StringBuilder(description).append("\n").append("Endpoint handler details:\n").append("Method [")
                                                .append(handlerMethod.getMethod()).append("]\n").append("Bean [")
                                                .append(handlerMethod.getBean()).append("]\n");
        return sb.toString();
    }
\end{minted}
\rule{\textwidth}{0.4pt}
\begin{minted}[fontsize=\footnotesize, linenos, firstnumber=56]{java}
public InvocableHandlerMethod getHandlerMethod() {
    return handlerMethod;
}
\end{minted}
\caption{An example of an Infer warning where non-local context does not provide enough information about the value of a class field. Infer warns that \mintinline{java}{handlerMethod} can be null (top). \mintinline{java}{getHandlerMethod()} simply returns the \mintinline{java}{handlerMethod} class field (bottom). But this is not enough to determine the legitimacy of this warning.}
\label{figure:class_field_ex}
\end{figure*}

\subsubsection{GPT-C}
DeepInferEnhance is able to boost vanilla Infer's precision, but requires labeled data. Since this data is expensive to collect, we sought a solution that could forego supervised learning altogether. Transformer models such as CodeBERT have shown promising performance in zero-shot settings for source code. Our approach is novel due to our interpretation of code completion recommendations by self-supervised generative models: recommendations for null checks are a signal that the null dereference warning is legitimate. With this approach, GPT-C had the highest precision of our models, improving on Infer by relative 17.5\%, with  slightly lower recall than DeepInferEnhance.

While analyzing the results from GPT-C, we identified several patterns in the warnings that GPT-C predicts incorrectly. These patterns included: 

\textbf{Insufficient or nonexistent non-local context} Our investigation revealed that when non-local context is unavailable or contains insufficient information, GPT-C does not perform well. For example, method calls belonging to an interface type cannot be resolved until runtime. Therefore, we cannot retrieve such methods as non-local context.
Similarly, non-local context can include getter methods that return a class field; however, these methods do not provide any information about the value the class field may hold. In the example in Figure \ref{figure:class_field_ex}, Infer warns that \mintinline{java}{handlerMethod}, which is assigned using \mintinline{java}{getHandlerMethod()}, can be null. \mintinline{java}{getHandlerMethod()} (middle) simply returns the \mintinline{java}{handlerMethod} class field (bottom). In order to correctly determine if the local variable \mintinline{java}{handlerMethod} can be null, we would need to collect not only the bodies of the methods \mintinline{java}{getHandlerMethod()} and \mintinline{java}{createMessagingErrorMessage()}, but also the constructors and fields of \mintinline{java}{DefaultAzureMessageHandler}.

\textbf{No reference to the target object} Null dereference warnings generally fall into two categories with respect to the target pointer. For some warnings, the null pointer is represented by a variable in the source code; for other warnings, the pointer is returned from a method with no explicit variable to hold its value. The latter case presents a problem for GPT-C recommendations: the nullable pointer has no reference in the code before the line where the warning occurs, which is not included in the input to the model. Therefore, this pointer does not appear in the input to GPT-C, reducing the chance that GPT-C recommends a null-check.

\textbf{Excessive sensitivity to the input} Several Infer warnings in a single project can refer to similar code, often with the same target variable or method. In such cases, if the instances are truly similar and legitimate, all of the instances should be reported as bugs to the end user. However, because GPT-C is very sensitive to minor differences in the input sequence, it may report only a subset of the warnings as legitimate. To enforce consistency, we group together warnings with the same target variable and label them all according to a logical OR, where all warnings are predicted as legitimate if GPT-C predicts any warning in the group as legitimate. The results in Table \ref{table:null_deref_results} include this consistency postprocessing. Alternatively, warnings could be grouped according to code similarity metrics such as edit distance.

\subsubsection{Overall Result}
DeepInferEnhance and GPT-C offer a tradeoff. Our objective is to increase precision, for which GPT-C is best. However, DeepInferEnhance has significantly better recall, capturing 88\% of legitimate bugs while still providing a 15\% boost in precision over vanilla Infer. Because of its superior recall, we would be more likely to recommend DeepInferEnhance to developers who value coverage in addition to precision. However, this model requires finetuning, whereas GPT-C offers the best precision and moderate recall without the need for additional data or further training.

\subsection{Experiment 2: Verifying the generalizability of neural approaches}\

To verify the generalizability of our neural approaches beyond the \emph{null pointer} bug, we evaluated our GPT-C model on Infer's \emph{resource leak} warnings. A resource leak happens when a program does not release resources it has acquired. The below code snippet shows an example of a resource leak warning, where an exception in \mintinline{java}{f.write(7)} will cause the program to skip the \mintinline{java}{f.close()} statement and leak the stream resource.
\begin{minted}[fontsize=\footnotesize, linenos, firstnumber=56]{java}
  public static void foo () throws IOException {
    FileOutputStream f = new FileOutputStream(new File("w"));
    f.write(7); //an exception here will cause a leak
    f.close();
  }
\end{minted}

In our target Java projects, Infer detected a total of 108 resource leak warnings. Table \ref{table:rl-repo-stats} shows the summary statistics of the identified resource leaks (Infer did not detect any resource leaks in Ambry).
\begin{table}[h!]
\centering 
 \caption{Summary statistics of resource leak warnings.Total warnings and true positives are as reported by vanillaInfer.}
 \label{table:rl-repo-stats}
\begin{tabular}{c c c c c} \hline
Name         & Lines of   & Total    & True      & Precision\\ 
             & Code       & Warnings & Positives \\
\hline\hline
Project A	 &     35,527 &	       6 &         2 & 33.3\% \\\hline
Project B    &	   66,346 &	      49 &        33 & 67.3\% \\\hline
Azure SDK    &	3,555,286 &	      33 &        16 & 48.5\% \\\hline
Playwright   &	   21,094 &	       2 &         2 & 100 \% \\\hline
Nacos        &	   62,443 &	       7 &         2 & 28.6\% \\\hline
Azure Maven  &	   23,995 &	      11 &         7 & 63.6\% \\
Plugins\\\hline
\textbf{Total} & \textbf{3,764,691} & \textbf{108} & \textbf{62} & \textbf{57.4\%} \\\hline
\end{tabular}
\end{table}

Since this data set was not large enough to meaningfully finetune DeepInferEnhance, we decided to only focus on our GPT-C model, where no further training or finetuning is required. We only had to adjust our prompting logic. For resource leaks, we use prefixes (first three characters) of the method names \mintinline{java}{close()} and \mintinline{java}{release()} as the prompts. If the leaked resource is assigned to a variable, we also use this variable name as a prompt. Table \ref{table:rl_results} shows the results of using GPT-C to remove false positives in resource leak warnings. 

As shown in the table, GPT-C can improve Infer's precision by 5.5\%. However, it fails to identify over a third of legitimate bugs. One pattern in the missed bugs is that some leaked resources have names or types, such as \mintinline{java}{EntityNotFoundHttpResponse} or \mintinline{java}{ChangeFeedProcessorBuilderImpl}, that do not clearly indicate that they are in fact resources, and therefore should be released. Types that clearly indicate resources, such as those that contain \mintinline{java}{File} or \mintinline{java}{Stream} in the name, are recognized more often by GPT-C.

\begin{table}[h!]
\centering 
 \caption{Performance of machine learning for removing false positive resource leak warnings}
 \label{table:rl_results}
\begin{tabular}{c c c c c} \hline
Approach      & Precision &  $\Delta$ Precision & Recall \\
\hline\hline
GPT-C         &   60.6\% &               5.56\% & 64.5\% \\
\hline
\end{tabular}
\end{table}

\section{Discussion}
Our GPT-C model improved Infer's precision by 17.5\% for \emph{null dereferences} and by 5.5\% for \emph{resource leaks}. However, it missed some of the correct warnings that Infer detected, with a recall of 84\% for \emph{null dereferences} and 65\% for \emph{resource leaks}. We identified several patterns of false negative predictions, which resulted in the reduced recall. One pattern occurred when the non-local context was a class field getter method. These methods are often a single return statement, which is not sufficient information for GPT-C to make the correct prediction. One way to mitigate this problem could be to include class fields and constructors as part of non-local context. However, the current GPT-C model is only trained for line completion using single method bodies. Newer transformer models for code, which use supplementary context in addition to individual method bodies, can better leverage this context to create more complete representations of the program state. Therefore, future work should explore training an extended-context model for code completion, as an evolution of the GPT-C model we used. We expect that such a model would perform better in many downstream tasks, including for verifying true positive warning from static analyzers.

Another class of warnings for which GPT-C did not perform well were chained method calls (e.g. \mintinline{java}{foo.bar().baz()}). If a warning is triggered on a method call in the middle of a chain, GPT-C cannot reasonably predict a null check. Since the intermediate method call is not stored in a variable, we cannot prompt GPT-C to predict a null check for the return value of that method call. One way to mitigate this problem is to modify the source code to insert a variable assignment for each method call in the chain. However, a developer would only break the method chain for a null check where necessary. Therefore, breaking the chain may create an abnormal code pattern that GPT-C will not recognize. Alternatively, the variable assignment could be inserted for only one method in the chain. For each method call in the chain, we could insert an assignment, generate recommendations using GPT-C, and select the recommendation with the highest confidence. However, we decided on a much simpler approach to mitigate chained method calls: simply trust Infer's decision and predict such warnings as legitimate bugs.

For any bug detection system, precision and recall have significantly different downstream impacts for developers. Low precision means that developers waste time analyzing many false positive warnings, while low recall means that some legitimate bugs are not identified. Static analyzers have typically favored coverage and recall over precision, with the objective of maximizing the number of reported legitimate bugs. However, in practice, low precision reduces developer adoption of analysis tools \cite{johnson-icse13} due to the time developers waste on investigating false positives. Prior work has found that developers mostly use analysis tools in their spare time and tend to fix warnings in short working sessions. Therefore, they are primarily driven by time constraints when addressing bugs identified by static analyzers \cite{quangdo2020developerneeds}. As a result, we chose to focus on precision rather than recall; we believe presenting developers with higher quality warnings will lead to bugs actually being addressed, rather than ignored due to a lack of confidence or time constraints. However in certain cases, where recall is more important, our models can be used to re-rank the warnings so that developers are presented with more true positives first. This allows us to present all warnings to developers while prioritizing likely legitimate bugs.

We demonstrated the effectiveness of transformer models for two bug types and one tool, but we believe this approach should generalize to other languages and tools. Our experiment on resource leaks provides evidence of this. In addition, our GPT-C approach is not tied to a particular programming language, static analyzer, or warning type. Because we use GPT-C in a zero-shot setting, no further training is required. Customization may instead be required through unique ways of prompting GPT-C and specific signals to seek in its outputs; adjusting prompts should be the only change necessary to apply GPT-C to new bug types. For example, one way to apply this technique to buffer overflow bugs in C/C++ could be to search for bounds checking recommendations. Even DeepInferEnhance, which requires labeled data for finetuning, can be expanded to additional programming languages by pretraining on a larger and more diverse corpus.

In this work we applied large-scale transformers to further verify bugs that have already been localized by static analyzers. While this means that our approach will not find bugs beyond those reported by the static analyzers, it is a cost-effective way to leverage transformers for this problem. Large-scale transformer models are expensive to train and evaluate, and using them to scan every method or every line of a project can be prohibitively expensive. By applying these models to resolve warnings that have already been localized by a static analyzer, we ensure that transformers are utilized in a cost-effective way. However, there may be other ways to localize bugs and use transformers. For example, one can leverage prior work on bug localization to determine buggy files \cite{wang2016icse} and only examine those files as opposed to the entire program.

\subsection{Threats to Validity}

\subsubsection{Dataset Size}
Static analysis warnings are time-intensive to triage, since each warning requires a detailed review of the source code involved in the warning. It is expensive to collect a large dataset of labeled warnings, which is preferred when training transformer models. This is particularly impactful for DeepInferEnhance, which requires labeled data for both finetuning and evaluation. Although we used cross validation to compensate for the limited dataset size, all of our approaches would benefit from a larger dataset. In order to scale the dataset, we must present warnings to project owners for review. Through developer engagements, we found that this raises a cold-start problem: in order to receive appropriate attention and high-quality feedback, warnings must have sufficiently high precision, or else developers may not engage with warnings shared for labeling purposes. We believe that the precision improvements of the approaches discussed here serve as a solution to this cold-start problem, and will allow us to share warnings with a wider set of projects to scale our dataset.

\subsubsection{Evidence of Generalizability}
Our approach of using machine learning to augment and complement static analysis is designed generically to benefit any analyzer. However, in this study, we focus on one analyzer (Infer) and two categories of bugs (null dereference and resource leak) for one language (Java). To gain wider adoption among developers of diverse projects, our approach must demonstrate benefits across additional languages and bug types. Our experiments with resource leaks are our first attempt to demonstrate this. We plan to apply and evaluate our approach to C\#, as well as additional languages, as the next step for expanding our approach.

\subsection{Data Release}
Our dataset consists of warnings from Infer for various open source and proprietary software projects. Source code from proprietary projects was made available to us solely for research purposes. Since we do not own this data, we cannot release it publicly. We intend to release data from open source projects after we have worked with each project owner to resolve the issues, or otherwise verify with the owners the safety of releasing bug or vulnerability data.

\section{Conclusion}
Rule-based bug detectors and static analyzers have been widely adopted for detecting security vulnerabilities, functional bugs, and even performance issues. However, building an analyzer is non-trivial because of the difficulty of balancing precision and coverage: reporting only correct bugs and ensuring that all similar bugs are reported.

The majority of existing analyzers favor higher coverage to ensure completeness, and therefore they produce more false positive warnings. However, frequent false positive warnings are one of the main barriers to wider adoption of static analyzers in the software industry; this problem cannot be solved by the analyzers themselves. To close this gap, we augmented static analyzers with a variety of machine learning models. We experimented with both feature-based and neural models for false positive reduction. Our experiments on Infer, a well-known interprocedural static analyzer, showed that leveraging GPT-C in a zero-shot setting can improve the precision of null pointer warnings by 17.5\% and resource leak warnings by 6\%. 

One immediate direction for future work is to experiment with more warning types and languages to further verify the generalizability of our approach. Another direction involves training transformers with broader context. For instance, one may include the imports, constructors, class fields, and superclasses (in cases of inheritance) as part of the context while training. We expect this broader context to increase transformer effectiveness in general, and especially in zero-shot settings to augment other code analyzers. A third direction is to explore whether a generative transformer similar to GPT-C can be used in conjunction with a static analyzer to suggest fixes for some or all the bugs.

\bibliographystyle{ACM-Reference-Format}
\bibliography{references}

\end{document}